%% file: paper.tex
\documentclass[%
 aip,
 jcp,
 amsmath,amssymb,
 reprint,%
]{revtex4-1}

\usepackage[utf8]{inputenc}
\usepackage{amsmath}
\usepackage{graphicx}   
\usepackage{dcolumn}    
\usepackage{bm}         
\usepackage{hyperref}
\usepackage{color}
\usepackage{array}
\usepackage{siunitx}
\usepackage[normalem]{ulem}
\graphicspath{{./}{./Figures/}}

\let\jcp\undefined

\newcommand{\jcp}{J.\ Chem.\ Phys.\ }

\newcommand{\jctc}{J.\ Chem.\ Theory Comput.\ }
\newcommand{\jpca}{J.\ Phys.\ Chem.\ A }

\newcommand{\pnas}{Proc.\ Natl.\ Acad.\ Sci.\ }
\newcommand{\wn} {cm$^{-1}$}
\newcommand{\wns} {cm$^{-1}$\ }

\newcommand{\fig}[1]{Figure~\ref{#1}}
\newcommand{\sfig}[1]{Fig.~\ref{#1}} 

\newcommand{\ssec}[1]{Sec.~\ref{#1}} 

\newcommand{\eqn}[1]{Eq.~(\ref{#1})}

\newcommand{\gtnote}[1]{\textcolor{red}{#1}}
\newcommand{\gtsout}[1]{\textcolor{red}{\sout{#1}}}

\renewcommand{\gtnote}[1]{\textcolor{black}{#1}}
\renewcommand{\gtsout}[1]{}

\input{abbrev}

\begin{document}

\title{Mean-field Matsubara dynamics: analysis of path-integral curvature effects in rovibrational spectra
      }

\author{George Trenins}
\author{Stuart C.\ Althorpe}%
 \email{sca10@cam.ac.uk}
\affiliation{ 
Department of Chemistry, University of Cambridge, Lensfield Road,\\ Cambridge,
CB2 1EW, UK.%
}%

\date{\today}

\begin{abstract}
It was shown recently that smooth and continuous `Matsubara' phase-space loops  follow a quantum-Boltzmann-conserving {\em classical} dynamics when decoupled from non-smooth distributions, which was suggested as the reason that many dynamical observables appear to involve a mixture of classical dynamics and quantum Boltzmann statistics. Here we derive a mean-field version of this `Matsubara dynamics' which sufficiently mitigates its serious phase problem to permit numerical tests on a two-dimensional `champagne-bottle' model of a rotating OH bond. The Matsubara-dynamics rovibrational spectra are found to converge towards close agreement with the exact quantum results at all temperatures tested (200--800~K),  the only significant discrepancies being a temperature-independent 22 \wns blue-shift in the position of the vibrational peak, and a slight
broadening in its lineshape. These results are compared with centroid molecular dynamics (CMD) to assess the importance of non-centroid fluctuations. Above 250~K, only the lowest-frequency non-centroid modes are needed to correct small CMD red-shifts in the vibrational peak; below 250~K, more non-centroid modes are needed to correct large CMD red-shifts and broadening. The transition between these `shallow curvature' and `deep curvature' regimes happens when imaginary-time Feynman paths become able to lower their actions by cutting through the curved potential surface, giving rise to artificial instantons \gtnote{in CMD}.
 \end{abstract}
\maketitle

\section{\label{sec:intro}Introduction}

Results from a wide variety of approximate calculations
suggest that nuclear dynamics can often be treated classically when confined to a single Born-Oppenheimer surface, with most observable quantum effects originating in the quantum Boltzmann statistics. Well-known examples include reaction rates at low temperatures,\cite{billjpc,liurev,craig2,craig3,annurev,tstrev,tom1,tom2,iceland,kastner} where tunnelling is dominated by `instantonic' barrier statistics,\cite{iceland,kastner,billinst,jorsca,jorpap,tim} and the vibrational spectrum of liquid water.\cite{qtip4pf,paes1,paes2,trpmd2,liubill,liuwat,liurev,annurev,water_review,xantheas}

However, standard semi-classical theory\cite{hillery,billjpc,liurev} implies that such a `classical
dynamics--quantum statistics' regime does not exist, except at very short
times, since it predicts that classical dynamics does not conserve the quantum
Boltzmann distribution, and that real-time coherence is needed to keep systems
in thermal equilibrium. Practical simulation methods have been devised which
get round this apparent contradiction using heuristic
quantum-Boltzmann-conserving classical dynamics.\cite{caov1,craig1,trpmd1,trpmd3,liuprop} Centroid
molecular dynamics (CMD)\cite{caov1} and 
(thermostatted) ring-polymer molecular dynamics [(T)RPMD]\cite{craig1,trpmd1,trpmd3}  have proved to be especially practical,
\cite{annurev,tstrev,tom1,tom2,paes1,paes2,trpmd2,water_review}
 but the heuristic dynamics these methods employ works in some regimes and fails in others.\cite{marx1,marx2,jorsca,haber}

It was found recently\cite{mats1} that a semi-classical theory that combines classical dynamics
with quantum statistics can be derived if one assumes that the dynamics of the
smooth `Matsubara' components of the imaginary-time Feynman paths  becomes decoupled from the dynamics of the non-smooth
components.  Once such a decoupling is assumed, the dynamics of the smooth
components becomes classical, without further approximation (since the
effective $\hbar$ in the smooth space is zero). Also, the smoothness of the
paths ensures that the plain Newtonian dynamics\cite{plain} that they follow
conserves the quantum Boltzmann distribution by giving the paths a continuous
symmetry with respect to imaginary-time translation.

This `Matsubara'  dynamics is currently a hypothesis and it cannot be used as a practical method
because of a serious phase problem.\cite{mats1} However, comparison with the valid limits of
various heuristic methods suggests that Matsubara dynamics  does account correctly for the
emergence of classical dynamics at thermal equilibrium. For example, RPMD works
well for short-time properties such as reaction rates,\cite{craig2,craig3,tstrev,annurev,tom1,tom2,jorsca,tim} and is the short-time
limit of Matsubara dynamics;\cite{mats2} CMD works well when a mean-field description of
the quantum Boltzmann distribution is suitable,\cite{marx1,marx2} and is the mean-field average
of Matsubara dynamics (where the mean-field average is over all Matsubara modes
except for the centroids);\cite{mats2} the `planetary model' of Smith et al.\cite{smith1,smith2} works well
for high-frequency stretch modes in liquid water and is a locally harmonic
approximation to the Matsubara fluctuations around the centroid.\cite{planets} These
comparisons suggest that we should pursue Matsubara dynamics further, 
since it may lead to  better understanding and
improvement of practical methods such as CMD, (T)RPMD  and the planetary model.

Here, we strengthen the evidence that Matsubara dynamics gives the
correct theoretical description of classical dynamics and quantum Boltzmann
statistics. In doing so, we also obtain new insight into path-integral curvature effects in
vibrational spectroscopy, and why they cause problems for CMD.  The CMD method
works well for vibrational spectroscopy of water at ambient temperatures,\cite{paes1,paes2} but
breaks down at lower temperatures, giving red-shifts and distortions in the
spectral line shapes.\cite{marx1,marx2,trpmd2} However, the success of CMD at high temperatures gives us
a clue that the dynamical decoupling of the smooth modes from the non-smooth modes in
Matsubara dynamics (the origin of which was left unspecified in ref.~\onlinecite{mats1}) probably arises from mean-field averaging.

In \ssec{sec:theory}, we show that a mean-field formulation of Matsubara dynamics is
simpler to derive than the more general formulation of ref.~\onlinecite{mats1}.
The phase still makes the dynamics impractical as a method, but is sufficiently
tamed that Matsubara dynamics can be used to calculate the vibrational spectrum of a
two-dimensional `champagne-bottle' model of OH, as reported in \ssec{sec:spectra}. We 
find that including just the lowest frequency non-centroid modes 
corrects the CMD red-shift at  temperatures down to about 250~K, but that more modes need to be included
below this, where the CMD red-shift increases dramatically. This low-temperature breakdown is shown in \ssec{sec:instantons}
to result from the proximity of artificial centroid-constrained instantons, which form when the imaginary-time Feynman paths
 can lower their actions by cutting through the curved potential surface.
\ssec{sec:conclusions} concludes the article.

\section{Mean-field formulation of Matsubara dynamics%
\label{sec:theory}}

One way to obtain a mean-field formulation of Matsubara dynamics would be to mean-field average over the Matsubara Liouvillian, derived in ref.~\onlinecite{mats1}. However, it is illustrative to derive mean-field Matsubara dynamics from first principles, starting from 
the exact quantum Kubo-transformed time-correlation function
\begin{align}
C_{AB}(t) = \int_0^\beta\! {d\lambda\over \beta}\,\tr\!
            \left[
                    e^{-\lambda {\hat H}}\hat A e^{-(\beta-\lambda) {\hat H}}\etb \hat B \etf 
            \right] 
            \label{kubo}
\end{align}
where $\beta=1/k_{\rm B}T$ and $\hat H$ is the system Hamiltonian. To simplify
the algebra, we consider a one-dimensional system in which the operators $\hat
A$ and $\hat B$ are functions of position only; these results generalise easily
to  many dimensions and to operators involving momenta.

Following ref.~\onlinecite{mats1} and earlier work,\cite{shig, nandini, tim}
 we can re-write $C_{AB}(t)$ in `ring-polymer' form as
\begin{align}
        &C_{AB}(t) = \lim_{N\to\infty} \int d\bq \int d\bDelta \int d\bz \ A_N(\bq) B_N(\bz) \no\\
        & \times \prod_{l=1}^N \bra{q_{l-1} - \Delta_{l-1}/2} \ebN \ket{q_{l}+\Delta_l/2} \no\\
        & \times \bra{q_{l}+\Delta_l/2} \etf \kb{z_l} \etb \ket{q_{l} - \Delta_{l}/2} 
        \label{monkey}
\end{align}
where \gtnote{$\beta_N \equiv \beta/N$,} $\int d\bq\equiv \int_{-\infty}^\infty dq_1\dots\int_{-\infty}^\infty
dq_N$, and similarly for $\bDelta$ and $\bz$, and
\begin{align}
        \label{adef}
        A_N(\bq)=&{1\over N}\sum_{i=1}^NA(q_i)
\end{align}
and similarly for $B_N(\bq)$. 
Inserting complete sets of momentum states,
\begin{align}
        \label{moms}
        \delta(\Delta_l-\Delta'_{l})={1\over 2\pi\hbar}
        \int_{-\infty}^\infty dp_l \ e^{ip_l(\Delta_l-\Delta'_{l})/\hbar}
\end{align}
we obtain 
\begin{align}
         C_{ AB}(t) = \lim_{N\to\infty}
                    & {1\over(2\pi\hbar)^N}\int d\bq \int d\bp\, 
                      \left[ 
                              e^{-\beta{\hat H}}
                      \right]_{\overline N}({\bf p},{\bf q}) \no\\ 
                    &\times A_N(\bq)\,e^{\hat Lt} B_N(\bq)
         \label{wiggy}
\end{align}
where the generalized Wigner transform $\big[ e^{-\beta{\hat
H}}\big]_{\overline N}({\bf p},{\bf q})$ and the quantum Liouvillian $\hat L$
are given in the Appendix. We emphasise that no approximation has yet been
made; \eqn{wiggy} is just a generalization of the standard Wigner identity
which allows quantum time-correlation functions to be written in terms of
phase-space variables.\cite{hillery}

Following ref.~\onlinecite{mats1}, we introduce the free-ring-polymer
normal-mode coordinates\cite{odd,notation}
\begin{align}
        Q_n&={1\over N}\sum_{l=1}^NT_{ln}q_l, \quad n=0,\pm 1,\dots,\pm n_N
\end{align}
with $n_N=(N-1)/2$ and
\begin{align}
         T_{ln} = 
         \left\{
         \begin{array}{ll}
                  1 & n=0 \\
                  \sqrt{2} \sin(2\pi ln/N) & n=1,\dots,n_N \\
                  \sqrt{2} \cos(2\pi ln/N) & n=-1,\dots,-n_N
         \end{array}
         \right.\label{ttt}
\end{align}
and the associated frequencies 
\begin{align}
        \omega_n'={2\over \beta_N\hbar}\sin{\left(n\pi \over N\right)}
\end{align}
We then take the limit $N\to\infty$ and define the set of $M$ lowest frequency
modes ($|n|\le (M-1)/2$) to be the `Matsubara modes' ${\bf Q}_M$, so-called
because their associated frequencies simplify to 
\begin{align}
        \omega_n={2n\pi\over\beta\hbar} \label{eq:mats-freqs}
\end{align}
since $M\ll N$. The significance of the Matsubara modes is that any
linear combination of them gives a smooth and continuous distribution of $q$ as a
function of imaginary time.\cite{doll,markland,mats1} Inclusion of the other $|n|> (M-1)/2$
`non-Matsubara modes' gives, in general, a discontinuous non-differentiable distribution
in $q$, resembling a random walk. The Matsubara modes
${\bf P}_M$, ${\bf D}_M$  give similarly smooth distributions of $p$ and $\Delta$.

The {\em only} approximation we will make to the exact dynamics of \eqn{kubo}
is to assume that the quantum Louivillian operator $\hat L$ can be replaced by
its mean-field average
\begin{widetext}
\begin{align}
        \label{lmf}
        {\hat L}_{\rm MF}({\bf Q}_M,{\bf P}_M)=
                 \lim_{N\to\infty}{\int\!d{\bf p}\int\!d{\bf q}\, 
                 \left[ 
                         e^{-\beta{\hat H}}
                 \right]_{\overline N}{\bf \delta}_M({\bf q},{\bf Q}_M)
                                      {\bf \delta}_M({\bf p},{\bf P}_M)\,
                 \hat L({\bf p},{\bf q})
                 \over \int\!d{\bf p}\int\!d{\bf q}\, 
                 \left[ 
                         e^{-\beta{\hat H}}
                 \right]_{\overline N}{\bf \delta}_M({\bf q},{\bf Q}_M)
                                      {\bf \delta}_M({\bf p},{\bf P}_M)}
\end{align}
\end{widetext}
where
\begin{align}
        {\bf \delta}_M({\bf q},{\bf Q}_M)=
                \prod_{n=-n_M}^{n_M}\delta\!
                \left(
                        Q_n-{1\over N}\sum_{i=1}^NT_{in}q_{\gtnote{i}}
                \right)
\end{align}
is a product of Dirac $\delta$-functions in the Matsubara modes ${\bf Q}_M$,
${\bf \delta}_M({\bf p},{\bf P}_M)$ is similarly defined for  ${\bf P}_M$, and
$n_M=(M-1)/2$. We also need to expand $A_N(\bq)$ in terms of normal modes, then
truncate at $|n|\le n_M$, giving 
\begin{align}
        A_M({\bf Q}_M)=\lim_{N\to\infty}{1\over N}\sum_{i=1}^NA(\tilde q_i)
\end{align}
where
\begin{align}
        \tilde q_i=\sum_{n=-n_M}^{n_M}T_{in}Q_n
\end{align}
and similarly for $B_N(\bq)$. This last step can be justified by noting that
the ring-polymer distribution will damp off functions of $Q_n$ for sufficiently
large $n$, allowing $M$ in $A_M({\bf Q}_M)$ and $B_M({\bf Q}_M)$ to be treated
as a convergence parameter. However, we give no justification at present for
the use of \eqn{lmf},
except for the numerical results presented in Sec.~III.\gtnote{\cite{mori}}

On evaluating the mean-field average in \eqn{lmf} (see the Appendix), we find that
\begin{align}
        {\hat L}_{\rm MF}({\bf Q}_M,{\bf P}_M)=\!
                \sum_{n=-n_M}^{n_M}
                {P_n\over m}{\partial \over \partial Q_n}\!-\!
                {\partial {\cal F}({\bf Q}_M)\over \partial Q_n} {\partial \over \partial P_n} 
\end{align}
where ${\cal F}({\bf Q}_M)$ is the free energy
\begin{align}
        \label{free}
        e^{-\beta {\cal F}({\bf Q}_M)}
            &=\lim_{N\to\infty}
                \left(m\over 2\pi\beta_N\hbar^2\right)^{(N-M)/2}N^{M/2}\no\\
            &\times\int\!d{\bf q}\,
                e^{-\beta [W_N({\bf q})-S_M({\bf Q}_M)]}{\bf \delta}_M({\bf q},{\bf Q}_M)
\end{align}
in which $W_N({\bf q})$ is the  ring-polymer potential energy
\begin{align}
        W_N({\bf q})=V_N({\bf q})+
        {1\over N}\sum_{l=1}^N{m(q_{l+1}-q_l)^2\over 2(\beta_N\hbar)^2} 
\end{align}
where $q_{l+N}\equiv q_l$, $V_N({\bf q})$ is defined analogously to $A_N({\bf q})$, and
\begin{align}
        S_M({\bf Q}_M)={m\over 2}\!\!\sum_{n=-n_M}^{n_M}\omega_n^2Q_n^2
\end{align}
is the Matsubara component of the `polymer springs'. Taking the mean-field average over
the non-Matsubara modes has therefore made the dynamics classical.\cite{centroid} This is because the Matsubara
phase-space $({\bf P}_M,{\bf Q}_M)$ has an effective
Planck's constant of zero, as first noted in ref.~\onlinecite{mats1}.

Having made the mean-field approximation, we can integrate out the
non-Matsubara modes from the time-correlation function (see the Appendix),
obtaining
\begin{subequations}
\label{final}
\begin{align}
        c_{AB}(t) =
            &{1\over (2\pi\hbar)^M}\int\! 
            d{\bf P}_M\int\! d{\bf Q}_M\, e^{-\beta
            \left[ 
                {\bf P}_M^2/2m+{\cal F}({\bf Q}_M)
            \right]}\no\\
            &\times e^{i\beta\theta_M({\bf P}_M,{\bf Q}_M)}
                    A({\bf Q}_M)e^{\hat L_{\rm MF}t}B({\bf Q}_M)
\end{align}
where
\begin{align}
        \theta_M({\bf P}_M,{\bf Q}_M)=\sum_{n=-n_M}^{n_M}\omega_nP_nQ_{-n}
\end{align}
\end{subequations}
is the Matsubara phase. Following similar arguments to ref.~\onlinecite{mats1},
one can prove that $\theta_M({\bf P}_M,{\bf Q}_M)$ is a constant of the motion,
ensuring that ${\hat L}_{\rm MF}({\bf Q}_M,{\bf P}_M)$ conserves the quantum
Boltzmann distribution in \eqn{final}. At $t=0$, one may analytically continue
$P_n\to P_n+i\omega_nQ_{-n}$,\cite{mats2} which removes the phase and cancels out
$-S_M({\bf Q}_M)$  in \eqn{free}, leaving the (standard) ring-polymer
distribution.

Equations (\ref{lmf}) and (\ref{final}) give the mean-field version of
Matsubara dynamics. For $M=1$, they reduce to centroid molecular dynamics
(CMD);\cite{caov1} for $M>1$, they generalise the dynamics to include $M-1$
non-centroid Matsubara modes. As mentioned above, the mean-field averaging in
\eqn{lmf} is the {\em only}  approximation made to the exact quantum dynamics;
we make no attempt here to justify it, but report numerical comparisons with
the exact quantum results in the next Section. 

\section{Matsubara dynamics of a vibrating-rotating OH bond%
\label{sec:spectra}}
\subsection{Two-dimensional `champagne-bottle' model}

We applied the mean-field Matsubara equations \eqn{final} to a two-dimensional
`champagne-bottle' model of a vibrating and rotating OH bond, similar to that
used in refs.~\onlinecite{marx1,marx2}. The radial polar coordinate $r$ represents the OH bond
length and the polar angle $\theta$ represents rotation in a plane. The
potential is taken to be a Morse function
\begin{align}
\label{eq:morse}
        V(r)=D_0 \left[
                1 - e^{-\alpha (r - r_{\mathrm{eq}})}
                 \right]^{\gtnote{2}}
\end{align}
with $r_\mathrm{eq} = 1.8324$,
$D_0 = 0.18748$ and $\alpha = 1.1605$ a.u.; the reduced mass $\mu=1741.05198$ a.u.
 The absorption intensity is calculated as
\begin{align}
        \label{eq:golden-rule}
        n(\omega) \alpha(\omega) \propto {1 \over 2\pi Z}\int_{-\infty}^\infty\!dt\, e^{-i\omega t}
        C_{\dot{\mu}\dot{\mu}}(t) f(t)
\end{align}
where $Z$ is the quantum partition function, $C_{\dot{\mu}\dot{\mu}}(t)$ is the
Kubo-transformed dipole-derivative autocorrelation function, and $f(t)$ is the
window function
\begin{align}
        f(t)=\dfrac{1}{
                1 + e^{(|t| - t_{1/2})/\tau}
                }
\end{align}
with parameters  $t_{1/2} = 400~\si{fs}$, $\tau = 25~\si{fs}$,  chosen 
to model the decorrelation time in liquid water.\cite{haber,liubill}
A linear dipole moment surface $\dot{\mu} = \dot{q}$ is used, with  the
proportionality constant in~\eqn{eq:golden-rule} set to unity. 

\fig{fig:eps1} plots the exact quantum spectrum (calculated using a discrete variable
representation) at  200--800~K. These temperatures are sufficiently low with
respect to the vibrational spacing (3590 cm$^{-1}$) that the centre of the
vibrational peak is temperature-independent.

\subsection{CMD calculations}

\begin{figure}[h]
\includegraphics{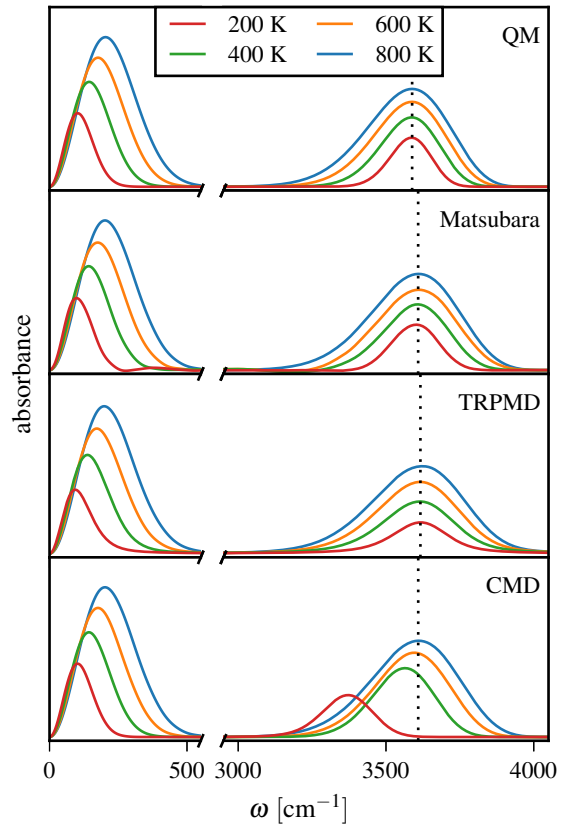}
\caption{\label{fig:eps1} 
        Mean-field Matsubara simulations of the two-dimensional champagne-bottle rovibrational spectrum, compared with the exact quantum results, and with TRPMD and CMD. Note the temperature-independent absorption maxima of the Matsubara and TRPMD vibrational peaks and the CMD red-shift which grows rapidly on decreasing the temperature from 400 to 200~K. }
\end{figure}

For $M=1$, the mean-field Matsubara equations \eqn{final} are equivalent to CMD. We used standard PIMD methodology\cite{pariram,annurev,hone,pile,chanwol,ceperley,tuckerman} to 
calculate the CMD approximations to the vibrational spectrum of the champagne-bottle model. 
The mean-field forces were evaluated on a regular grid, using cubic spline interpolation
to approximate the intermediate values. Mean-field force calculations
were performed with 64, 32, and 16 beads 
at 200, 400, and 600--800~\si{K}, on a grid of 64 points from 0.5--2.0~\si{\angstrom} at 400--800~\si{\K}, 
and 128 points at 200~\si{\K}.

The results of the  CMD calculations are shown in \sfig{fig:eps1}, and
exhibit the well-known `curvature problem',\cite{marx1,marx2}  whereby the CMD vibrational peak
shifts to the red as the temperature is lowered. Two aspects of this behaviour
are worth pointing out. First, at 800~K, the CMD peak is in very close
agreement with the exact quantum peak, except for a small blue-shift (22~\wn)
and a slight overestimate in the width of the peak. Note that the classical peak at this temperature (not shown) is
blue-shifted by about 105~\wns on account of zero-point energy violation, and is
similarly broadened. Second, the red-shifting of the CMD peak increases gradually down to about
250~K, and the line-shape scarcely changes; but below about 250~K, the
red-shift increases dramatically (to 215~\wns at 200~K),
and the line-shape broadens
noticeably. We return to these two points below.

\subsection{Mean-field Matsubara calculations}

Mean-field Matsubara spectra for $M>1$ were calculated using a straightforward
generalization of \eqn{final} to  $2M$ Matsubara modes $({\bf X}_M,{\bf Y}_M)$,
with $A_M=B_M$ taken to be $\dot X_0$ and $\dot Y_0$.
For $M>1$, it is only practical to evaluate  ${\cal F}({\bf X}_M,{\bf Y}_M)$ on the
fly, using an extension to $2M${~}modes of the partially-adiabatic CMD technique
of ref.~\onlinecite{hone}. This entails using $N$-bead ring-polymers, with
the mean-fielding over the $2(N-M)$ highest modes accomplished through adiabatic
decoupling, by shifting the respective frequencies to a large value $\Omega$
and re-scaling the associated masses $m_n = m (\omega_n / \Omega)^2$. To ensure
proper sampling, a Langevin thermostat is attached to each of the mean-fielded 
modes, with the friction coefficient set to the optimal value of $2 \Omega$.\cite{pile}

Converged $M=3$ spectra were obtained for $N=32$ at 200~\si{K} and $N=16$ at 400--600~\si{K};
the $M=5$ spectrum was calculated for $N=24$. The adiabatic frequency
was taken to be $\Omega = \Gamma / \beta_N \hbar$, with the adiabatic separation
$\Gamma = 32$ at all temperatures. 
The drawback of this approach is that a small time-step $\Delta t$ is
needed to cope with the rapid motion of the mean-fielded modes; we used $\Delta
t=0.003125~\si{fs}$. 

As expected, the most challenging part of the calculation was integrating
over the phase $\theta_M$, which was done by evaluating the ratio 
\begin{flalign}%
\label{eq:tcf-ratio}
        c_{AB}(t) = \dfrac{
                \left\langle
                \cos\left(\beta \theta_M\right) 
                A\left(\mathbf{Q}_M\right) e^{\hat{L}_\mathrm{MF} t} 
                B\left(\mathbf{Q}_M\right)
                \right\rangle
                          }{
                \left\langle \vphantom{e^{\hat{L}_\mathrm{MF} t}}
                \exp\left( -\beta \sum m \omega_n^2 Q_n^2 / 2 \right)
                \right\rangle
                           }
\end{flalign}
where $\langle\, \cdot \, \rangle$ denotes thermal averaging according to the 
distribution $
        e^{-\beta [\mathbf{P}_M^2 / 2m + \mathcal{F}(\mathbf{Q}_M)]}
             $, 
and the sum in the denominator is over the $M$ non-mean-fielded modes.
The sampling was done by averaging over an ensemble of partially adiabatic trajectories, each
1000~fs long. \gtnote{For a given number of modes $M$, the convergence is slower at higher temperatures, as the system samples more of the phase-space, making the integrand in the numerator of~\eqn{eq:tcf-ratio} more oscillatory.} With the computing resources available, we were unable to go beyond $M=1$ at
800~K, $M=3$ at 600 and 400~K, and $M=5$ at 200~K. For the $M=3$ calculations, \num{6e6}, \num{3e7}, and \num{8e7} trajectories were used at 200, 400, and 600~\si{K};
for $M=5$, \num{3e8} trajectories were used\gtnote{, the latter taking three weeks on 128 CPU cores to complete.}
Even within these limits, small numerical artifacts are likely to remain in the rovibrational spectra,
resulting from imperfect adiabatic separation and sampling.

\begin{figure}
\includegraphics{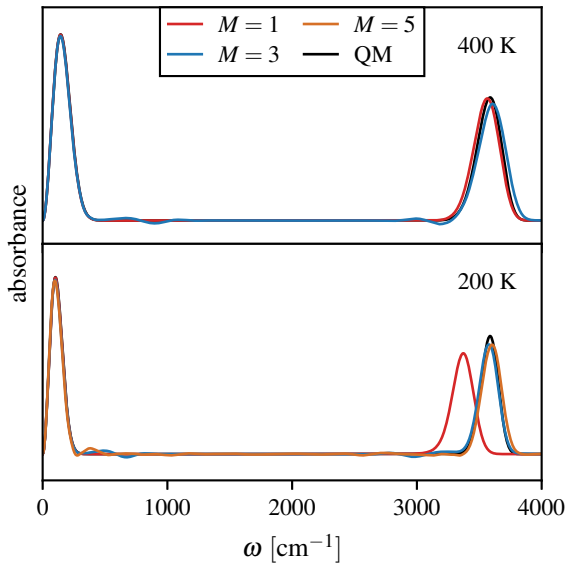}
\caption{\label{fig:eps2} 
Convergence of the mean-field Matsubara rovibrational spectrum with respect to the number of non-mean-fielded modes $2M$. 
Spectra for $M = 3$ and $M = 5$ at $400$ and $200~\si{K}$ respectively are also plotted in \sfig{fig:eps1}.
}
\end{figure}

\fig{fig:eps2} illustrates the convergence of the mean-field Matsubara results with
respect to $M$. As mentioned above, we were unable to include more than a few non-mean-fielded modes, owing to the oscillatory
Matsubara phase. However, the results for $M=1,\,3,\,5$ at 200~K (\sfig{fig:eps2}), for
which the CMD red-shift is greatest, suggest that these small values of $M$ are
sufficient to converge the position and the overall shape of the vibrational
peaks. Some convergence artifacts remain, visible as 
`wiggles' in the spectra in \sfig{fig:eps2}. These artifacts are not sampling errors: they are the result of incomplete
convergence with respect to $M$ and indicate that a small component of the dynamics requires a long `tail' of Matsubara modes to be described correctly. Some of the wiggles can be made to disappear if the fluctuations around the centroid are approximated by local normal modes (these results not shown), suggesting that they are caused by vibration-rotation coupling. The convergence `tail' is thus probably the result of using cartesian rather than polar Matsubara modes.

Even with the convergence errors discussed above, the Matsubara results in
\sfig{fig:eps1} are in strikingly good agreement with the exact quantum results, across
the entire 200--800~K temperature-range tested. Most importantly, the Matsubara
vibrational peak positions are correctly independent of temperature, with the
22~\wns blue-shift observed in the CMD results at 800~K remaining constant down
to 200~K to within the sampling error.\cite{serrors} The slight broadening of the
vibrational line-shape seen in the CMD results at 800~K also continues in the
Matsubara results down to 200~K (although the line-shapes are likely to be
somewhat distorted by the convergence errors mentioned above). 
 If we rule out the possibility
of a long convergence tail in $M$ changing the position of the vibrational
peak,  we can infer that the 22~\wns red-shift and the slight narrowing of the quantum vibrational
peak with respect to the Matsubara peak are the only significant real-time coherence effects. 

Subject to these caveats, we can also infer that CMD agrees closely with Matsubara dynamics at 800~K, and gives a reasonable approximation to it down to about 250~K. In this temperature range, the CMD red-shifts are small and can be corrected by including just the $|n|$=1 Matsubara modes. 
 However, below 250~K, the CMD red-shift increases dramatically.  At 200~K, the $|n|=2$ modes  are also needed to correct the red-shift, and many more modes would be required at lower temperatures.

\section{%
\label{sec:instantons}%
Centroid-constrained instantons}

\begin{figure}
\includegraphics{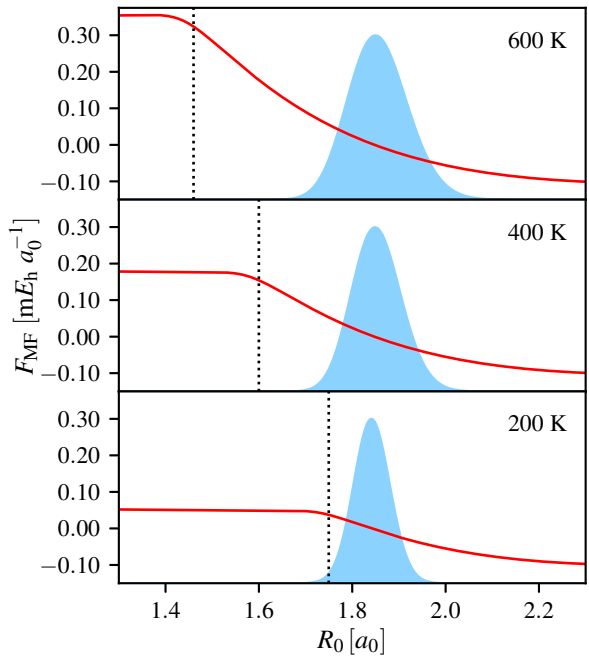}
\caption{\label{fig:eps3}%
The CMD mean-field force $-d\mathcal{F}(R_0)/dR_0$ (red line) plotted on top of the corresponding Boltzmann distribution \mbox{$\propto R_0 e^{-\beta \mathcal{F}(R_0)}$} (shaded blue). The dotted vertical lines indicate the position of the critical radius $R_c$ given by \eqn{eq:rc}. Note that $R_c$ coincides with the onset of the flattening of the force, and that the Boltzman distribution overlaps $R_c$ at 
$200~\si{K}$.}
\end{figure}

To investigate why CMD breaks down rapidly below 250~K, we plot in \sfig{fig:eps3} the centroid mean-field force 
 $-d{\cal F}/d R_0, R_0=\sqrt{X_0^2+Y_0^2}$, at 200--600~K, and
overlay this with the CMD Boltzmann distribution as a function of $R_0$. As has been
noted previously,\cite{marx1,marx2} the force flattens out for values of $R_0$ less than a
certain radius, and this radius increases as the temperature decreases. \fig{fig:eps3} shows immediately why CMD breaks down below about 250~K: at 400 and 600~K,  the quantum Boltzmann distribution is well separated from the flat
region, but at 200~K, the distribution starts to overlap it.

\begin{figure}
\includegraphics{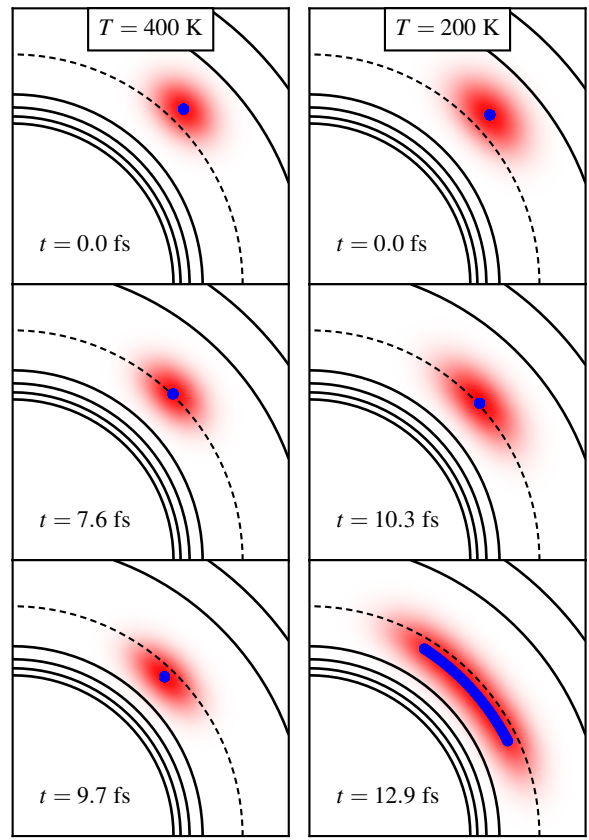}%
\caption{\label{fig:eps4}%
Snapshots of CMD trajectories on the Morse potential of \eqn{eq:morse} (black contour lines, $r_{\mathrm{eq}}$ dotted), with centroid-constrained bead distributions shown in red and corresponding minimum-energy ring-polymer configurations in blue.  Note the artificial instanton in the 200~K trajectory at 12.9~fs.}
\end{figure}

It is easy to identify the origin of the flattening. \fig{fig:eps4} shows the centroid-constrained
ring-polymer distribution at three points along a single trajectory at 400~K,
and at 200~K. The 200~K trajectory is one of the 6$\%$ of trajectories that
make it into the flat region at this temperature. During the 400~K trajectory,  the
distribution moves as a relatively compact `blob', stretching slightly at the
inner turning point as it pushes against the repulsive wall; the minimum-energy
ring-polymer  within the distribution (i.e.~the imaginary-time Feynman path with the least action)
is a point at
the centroid.  During the 200~K trajectory, by contrast, the distribution smears out at the turning
point,  where the  minimum-energy ring-polymer has a delocalised
geometry (\sfig{fig:eps4}). Since this geometry is an extremal point
on the ring-polymer surface, subject to the centroid constraint, the path
followed by the beads corresponds to a periodic orbit on the inverted potential
surface, subject to a time-averaged constraint. In other words, by constraining the centroid in the distribution, the
CMD method creates artificial instantons below 250~K.

\begin{figure}[b]
\includegraphics{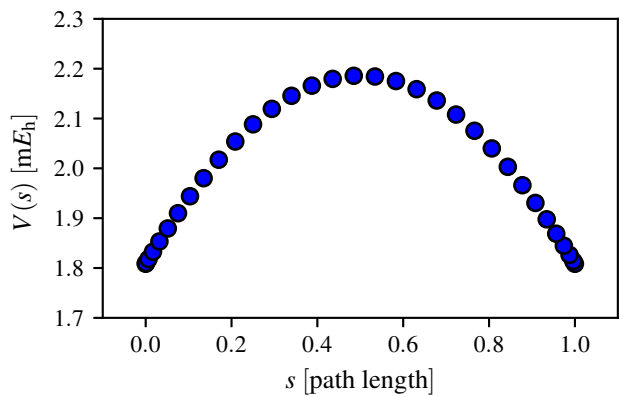}%
\caption{\label{fig:eps5}%
Potential energy along the beads (discrete imaginary-time steps) of the artificial centroid-constrained instanton shown in \sfig{fig:eps4}. The path length $s$ is taken to be a linear function of the polar coordinate $\theta$. The potential energy varies because the path followed by the instanton in \sfig{fig:eps4} is gently parabolic. 
}
\end{figure}

We can make analogies with instanton formation in quantum rate theory\cite{billinst,jorsca,jorpap,iceland,kastner} to
understand what is happening at these lower temperatures. In rate theory, instantons form below a
cross-over temperature; in the CMD dynamics considered here, it is more
convenient to define a `cross-over radius' $R_c$. By minimising the ring-polymer energy subject to the centroid constraint, one
can show (see the supplementary material) that
\begin{align}
\label{eq:rc}
R_c \simeq \left. -\dfrac{1}{m \omega_1^2} \dfrac{d V}{d r}\right|_{r=R_c}
\end{align}
\gtnote{where $\omega_1$ is the first Matsubara frequency as defined in \eqn{eq:mats-freqs}.}
The values of $R_c$ at 200--800~K are shown in \sfig{fig:eps3}, and are found to coincide with the onset of the flat region of the centroid force.
For $R_0<R_c$, the potential
is sufficiently curved that a centroid-constrained ring-polymer can minimize
its energy by stretching and moving outwards (leaving the position of the centroid unchanged); it cannot stretch around a perfectly circular path, since this would correspond to a purely rotational periodic orbit on the inverted potential, with a period greater than $\beta\hbar$; so the orbit follows a gently parabolic curve which cuts  through the circular potential energy surface. The variation of $V$ along the parabolic curve is plotted in \sfig{fig:eps5}, which shows  that the imaginary-time periodic orbit on the inverted potential resembles a conventional instanton or `bounce' in barrier tunnelling.\cite{billinst,jorsca,iceland,kastner} For
$R_0>R_c$, the potential is not sufficiently curved for the ring-polymers to be able to lower their energy
by cutting through the potential, hence the minimum-energy ring-polymer collapses to a point at the centroid.

The two temperature regimes are thus analogous to the `shallow' and `deep' tunnelling regimes in reaction rate theory,\cite{jorsca} with 250~K being the approximate `cross-over temperature' for the OH model. Just as in rate theory, the notion of a precise cross-over temperature is somewhat artificial, since it refers to the switch in the position of the ring-polymer stationary point from the collapsed to the instanton geometry. In rate-theory, instanton-like delocalisation starts to happen above cross-over, as a result of softening of the lowest-frequency Matsubara mode. Analogous behaviour is responsible for the large red-shift in the CMD vibrational peak at 200~K. Only 6$\%$ of the CMD trajectories make it into the flat region ($R_0<R_c$), but a majority of trajectories get sufficiently close to $R_0=R_c$ for the first Matsubara mode to soften appreciably. At lower temperatures (not shown here) all the CMD trajectories enter the flat region to form instantons, giving rise to much greater red-shifts and broadening of the vibrational peak (e.g.~see the 100~K red-shifts calculated for a similar OH model in ref.~\onlinecite{marx1}).

It is important not to push  the analogy with rate theory too far: the instantons in rate-theory are real, but the centroid-constrained instantons identified above are artificial.  However, the change in the quantum statistics that takes place at about 250~K is real: below this temperature, the ring-polymers are sufficiently floppy that they can lower their energy by cutting through the curvature of the potential surface. \cite{inapp}

\begin{figure}[b]
\includegraphics{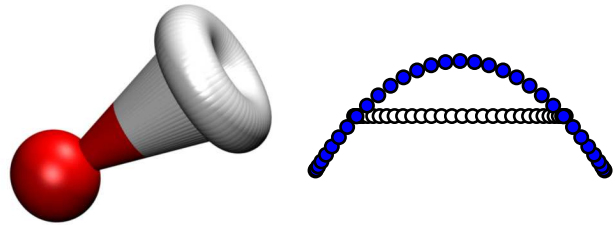}%
\caption{\label{fig:eps6}%
A circular-orbit artificial instanton formed during a CMD trajectory at 200~K \gtnote{in }a three-dimensional champagne-bottle model of OH (the instanton beads are shown as white spheres for the H-atom, red for the O-atom). Plotted on the right is a side-on view of the same instanton (in white), together with a parabolic instanton (in blue) from a two-dimensional calculation at the same temperature with the same centroid constraint. 
}
\end{figure}

Periodic orbits and related objects can sometimes show special behaviour in
two dimensions (2D), and for this reason we also examined centroid trajectories in the
three-dimensional (3D) version of the model. \cite{3diff} We found that the extra degree
of freedom permitted a different type of  instanton to form, corresponding to a
circular periodic orbit in a plane tangential to $R_0$ on the inverted
potential surface. One of these circular instantons is shown in \sfig{fig:eps6}.
However, the 3D centroid-constrained distributions behave very similarly to the
2D distributions, because the cross-over radius $R_c$ for the 3D circular orbits is the same
as for the 2D parabolic orbits, which also extremise the action in 3D (see the supplementary material).
 As a result, the 3D mean-field centroid force
flattens out at the same radial displacement as the 2D force. Curvature effects are slightly bigger in 3D 
because the Boltzmann distributions overlap the flat region slightly
more.  Similar circular instantons
have also been found in CMD distributions for gas-phase water,\cite{andreea}
suggesting that the 2D picture developed here applies to vibrational
spectroscopy generally.

\section{%
\label{sec:conclusions}%
Conclusions}

We have shown that Matsubara dynamics can be derived more simply as a
mean-field theory. This does not solve the phase problem, but does make the approach sufficiently practical to treat model systems.
In tests on a two-dimensional model of a rotating OH bond, the  Matsubara vibrational spectra were found to agree closely with the exact quantum results over the entire 200--800~K temperature range tested. This is a strong piece of evidence in support of the idea that Matsubara dynamics accounts for the classical part of the exact dynamics in a quantum Boltzmann distribution.
Real-time quantum
coherence effects were found to be minor in the OH model: a 22~\wns red-shift
in the position of the quantum vibrational peak (with respect to the Matsubara result), and a slight narrowing in its shape.
It seems reasonable to expect a comparably small red-shift and narrowing in the  OH-stretch band of bulk water.

We  also found that quantum Boltzmann statistics responds to the curvature of the OH potential in two distinct ways, giving rise to `shallow curvature' and `deep curvature' regimes which are loosely analogous to the `shallow tunnelling' and `deep tunnelling' regimes in quantum rate theory.\cite{jorsca} The cross-over temperature (250~K in the OH model) marks the point at which  imaginary-time Feynman paths can lower their actions by cutting through the curved potential surface. This behaviour gives rise to artificial instantons in CMD, explaining why CMD gives a reasonable approximation to Matsubara dynamics above the cross-over temperature, but a poor one below it. Although tested on a simple model, we expect this result to generalise, and for it to be possible to estimate the cross-over temperature in bulk systems by searching for centroid-constrained instantons that minimise the action. It is likely that the cross-over temperature for  the OH-stretch band in bulk water  is
 below freezing, since CMD works well for the liquid\cite{paes1,paes2} but gives significant vibrational red-shifts for ice.\cite{trpmd2}

\section*{Supplementary Material}

See supplementary material for a derivation of \eqn{eq:rc} for both the parabolic and circular instantons.

\begin{acknowledgments}
G.T.\ acknowledges a University of Cambridge Vice-Chancellor's award and
support from St.\ Catharine's College, Cambridge. 
S.C.A.\ acknowledges funding from the UK Science and Engineering Research
Council.
\end{acknowledgments}

\appendix
\section*{Appendix: Mathematical details}

The generalised Wigner transforms in \eqn{wiggy} are
\begin{widetext}
\begin{align}
\left[ e^{-\beta{\hat H}} \right]_{\overline N} ({\bf p},{\bf q})= & \int\! d\bDelta  \,
 \prod_{l=1}^N \bra{q_{l-1} - \Delta_{l-1}/2} \ebN \ket{q_{l}+\Delta_l/2} e^{ip_l\Delta_l/\hbar}
\label{wig1}
\end{align}
and 
\begin{align}
\left[ {\hat B}(t)\right]_{N}({\bf p},{\bf q}) = & \int d\bDelta  \int d{\bf z}  \ B_N(\bz) \no\\
 & \times \prod_{l=1}^N \bra{q_{l}-\Delta_l/2} \etf \kb{z_l} \etb \ket{q_{l}+ \Delta_{l}/2} e^{ip_l\Delta_l/\hbar}
 \label{wig2}
\end{align}
\end{widetext}
with $[\hat{B}(0)]_N(\mathbf{p}, \mathbf{q}) = B_N(\mathbf{q})$.

To obtain the quantum Liouvillian ${\hat L}_N$ in \eqn{wiggy}, we generalise the standard derivation of the Moyal series,\cite{hillery} differentiating $[ {\hat B}(t)]_{N}({\bf p},{\bf q})$
with respect to $t$, and using integration by parts to pull the Heisenberg time-derivatives in front of the integral, giving
\begin{align}
{d\over dt}\left[ {\hat B}(t)\right]_{N}({\bf p},{\bf q}) =  {\hat L}_N\left[ {\hat B}(t)\right]_{N}({\bf p},{\bf q})
\end{align}
where 
\begin{align}
 {\hat L}_N =\sum_{l=1}^N& \frac{p_l}{m} \frac{\partial}{\partial q_l} - V(q_l) \frac{2}{\hbar} \sin\!\left(\frac{\ola \partial}{\partial q_l}\frac{\hbar}{2}\frac{\ora \partial}{\partial p_l} \right)
\end{align}
and the arrows indicate that the differential operators act to the left and right respectively.

To evalute the mean-field integrals in \eqn{lmf}, we first rewrite ${\hat L}_N$ in terms of the normal-mode coordinates $({\bf P}_N,{\bf Q}_N)$ as
 \begin{align}
\lim_{N\to\infty}{\hat L}_N = {\cal L}_{M} +\lim_{N\to\infty} {\hat L}_{N,M} \label{things}
\end{align}
where 
\begin{align}
 {\cal L}_{M}& =\lim_{N\to\infty}\sum_{n=-n_M}^{n_M}{P_n\over m}{\partial \over \partial Q_n}\no\\&
  - V_N({\bf Q}) \frac{2N}{\hbar} \sin\!\left(\sum_{n=-n_M}^{n_M}   \frac{\hbar}{2N}{\ola \partial \over \partial Q_n}{\ora \partial \over \partial P_n} \right)\label{matlouf}
\end{align}
involves derivatives of only the Matsubara modes $({\bf P}_M,{\bf Q}_M)$, and
${\hat L}_{N,M}$  involves also derivatives of the non-Matsubara modes. We do
not need to know ${\hat L}_{N,M}$ explicitly (although it can easily be
obtained using trigonometric identities\cite{mats1,notation}), since its mean-field
average is zero on account of the derivatives in the non-Matsubara modes.
This leaves us with  ${\cal L}_{M}$, which simplifies (without approximation,
because $M\ll N$) to
\begin{align}
 {\cal L}_{M} =\lim_{N\to\infty}\sum_{n=-n_M}^{n_M}{P_n\over m}{\partial \over \partial Q_n}
  - {\partial V_N({\bf Q})\over \partial Q_n} {\partial \over \partial P_n} 
\end{align}
with $V_N({\bf Q})$ defined analogously to $A_N({\bf Q})$ of \eqn{adef}.

 To carry out the mean-field average in \eqn{lmf}, we therefore need to evaluate the integrals
 \begin{align}
I_n({\bf P}_M,{\bf Q}_M)=\int\!d{\bf p}\int\!d{\bf q}\, &\left[ e^{-\beta{\hat H}}\right]_{\overline N}{\bf \delta}_M({\bf q},{\bf Q}_M)\no\\&\times{\bf \delta}_M({\bf p},{\bf P}_M)\,{\partial V_N({\bf Q})\over \partial Q_n}
\end{align}
Integrating over $\bf p$ gives
\begin{align}
I_n({\bf P}_M,&{\bf Q}_M)=(2\pi\hbar)^{N-M}N^M\int\! d{\bf D}_M\int\! d{\bf q}\no\\\times&
\bra{q_{l-1}-\eta_{l-1}/2}e^{-\beta_N\hat H}\ket{q_{l}+\eta_{l}/2}\no\\
\times& {\bf \delta}_M({\bf q},{\bf Q}_M){\partial V_N({\bf Q})\over \partial Q_n}\prod_{k=-n_M}^{n_M}e^{iD_kP_kN/\hbar}
\end{align}
with
\begin{align}\label{eta}
\eta_{l}=\sum_{n=-n_M}^{n_M}T_{ln}D_n
\end{align}
Writing the bra-kets as 
\begin{align}
\lim_{N\to\infty}&\bra{q_{l-1}-\eta_{l-1}/2}e^{-\beta_N\hat H}\ket{q_{l}+\eta_{l}/2}=\no\\&\left(m\over 2\pi\beta_N\hbar^2\right)^{1/2}
e^{-\beta_N\left[V(q_l+\eta_l/2)+V(q_{l-1}-\eta_{l-1}/2)\right]/2}\no\\
&\times e^{-[q_l-q_{l-1}+(\eta_l+\eta_{l-1})/2]^2m/2\beta_N\hbar^2}
\end{align}
we obtain
\begin{align}
&\lim_{N\to\infty}I_n({\bf P}_M,{\bf Q}_M)=\lim_{N\to\infty}(2\pi\hbar)^{N-M}\left(m\over 2\pi\beta_N\hbar^2\right)^{N/2}\no\\&\,\times N^M\int\! d{\bf D}_M\int\! d{\bf q}\,{\bf \delta}_M({\bf q},{\bf Q}_M){\partial V_N({\bf Q})\over \partial Q_n}\no\\&\,\times
\prod_{l=1}^Ne^{-\beta_N[V(q_l+\eta_l/2)+V(q_l-\eta_l/2)]/2}e^{-(q_l-q_{l-1})^2m/2\beta_N\hbar^2}\no\\
&\,\times\! \prod_{k=-n_M}^{n_M}\! e^{-D_k^2N^2m/2\beta\hbar^2}e^{D_kQ_{-k}\omega_kNm/\hbar}e^{iD_kP_kN/\hbar}\end{align}
where we have made use of the orthogonality of ${\bf T}$, and the relations
\begin{align}
T_{l+1\,n}=&T_{ln}+{\cal{O}}(N^{-1})\no\\
T_{l+1\,n}-T_{l\,n}=&T_{l\,-n}\omega_n\beta_N\hbar+{\cal{O}}(N^{-2})
\end{align}
(easily proved using trigonometric identities). In the limit $N\to \infty$, the integrals over ${\bf D}_M$ can be done analytically (since the $e^{-D_k^2N^2m/2\beta\hbar^2}$ terms allow one to neglect the $\eta_l$-dependencies in $V$), giving
\begin{align}
\lim_{N\to\infty}\!I_n({\bf P}_M,&{\bf Q}_M)=\!\lim_{N\to\infty}\!N^{M/2}\!
\left(\!2\pi m\over \beta_N\!\right)^{\!(N-M)/2}\!\!\!\!
 e^{-\beta \mathbf{P}_M^2/2m}\no\\&\times
e^{i\beta\theta_M({\bf P}_M,{\bf Q}_M)}\int\! d{\bf q} \, e^{-\beta [W_N({\bf q})-S_M({\bf Q}_M)]}\no\\&\times{\bf \delta}_M({\bf q},{\bf Q}_M){\partial V_N({\bf Q})\over \partial Q_n}
\end{align}
Substituting this expression into \eqn{lmf}, and evaluating the analogous
integral in the denominator gives \eqn{free}. A similar integration over ${\bf
p}$ and ${\bf D}_M$ in the time-correlation function (noting that $A_M({\bf
Q}_M)$ and $e^{\hat L_{\rm MF}t}B_M({\bf Q}_M)$ are independent of the
non-Matsubara modes)  gives \eqn{final}.

\nocite{*}

\end{document}

%% file: abbrev.tex
\newcommand{\ebN}{e^{-\beta_N \hat H}}

\newcommand{\etf}{e^{-i \hat H t/\hbar}}
\newcommand{\etb}{e^{i \hat H t/\hbar}}

\newcommand{\bra}[1]{\langle #1 |}
\newcommand{\ket}[1]{| #1 \rangle}
\newcommand{\kb}[1]{\ket{#1}\bra{#1}}

\newcommand{\tr}{ {\rm Tr} }

\newcommand{\no}{\nonumber}

\newcommand{\ola}{\overleftarrow}
\newcommand{\ora}{\overrightarrow}

\newcommand{\bp}{ {\bf p} }

\newcommand{\bq}{ {\bf q} }

\newcommand{\bz}{ {\bf z} }
\newcommand{\bDelta}{ {\bf \Delta} }















%% file: paper.bbl
\begin{thebibliography}{10}
\bibitem{plain} By which we mean that the classical dynamics does not involve ring-polymer springs.
\bibitem{notation} Note that the definition of $Q_n$ used in ref.~\onlinecite{mats1} differs from that used here by a factor of $\sqrt{N}$.
\bibitem{centroid} This result was derived earlier for the special case of the centroid in ref.~\onlinecite{shig}.
\bibitem{serrors} The calculated blue-shifts are 22, 22, 18, and 14 \wns at 800, 600, 400, and 200~K, with the result at 200~K having an error bar of the order of 10 \wn.
\bibitem{inapp} In this sense the CMD instantons are analogous to the artificial instantons that sometimes form in quantum rate theory if an inappropriate dividing surface is used; see the discussion in Sec.~IIIC of ref.~\onlinecite{tim}.
\bibitem{3diff} The difficulty of converging the Matsubara calculations did not permit us to extend the two-dimensional analysis just reported to three dimensions. Preliminary calculations show that in addition to the phases being more oscillatory, the three-dimensional system requires more non-centroid modes because the rotation-vibration `wiggles' (see Sec.~IIIC) are more pronounced.
\end{thebibliography}
